\documentclass[useAMS,usenatbib,twocolumn]{mn2e}
\usepackage{natbib}
\usepackage{graphicx}
\usepackage{times}
\usepackage{rotate}
\usepackage{color,url}
\usepackage[dvipdfm,%
bookmarkstype=toc=true,%
linktocpage=true,%
bookmarks=true%
 ]{hyperref}

\newcommand{\prd}{Phys.~Rev.~D}
\newcommand{\aap}{A.\&Ap.}
\newcommand{\mnras}{MNRAS}
\newcommand{\apj}{ApJ}
\newcommand{\aj}{AJ}
\newcommand{\apjl}{ApJ}

\newcommand{\apjs}{ApJS}

\newcommand{\jcap}{JCAP}

\newcommand{\bm}[1]{\mbox{\boldmath$#1$}}
\newcommand{\simgt}{\lower.5ex\hbox{$\; \buildrel > \over \sim \;$}}
\newcommand{\simlt}{\lower.5ex\hbox{$\; \buildrel < \over \sim \;$}}
\newcommand{\ave}[1]{\left\langle #1\right\rangle}

\newcommand{\bvec}[1]{\mbox{\boldmath $#1$}}
\newcommand{\sqdeg}{{\rm deg}$^{-2}$}

\newcommand{\id}{{\rm d}}

\begin{document}
\title[BAO with the cross-correlation]
{
  Baryon acoustic oscillations with the cross-correlation of
  spectroscopic and photometric samples
}
\author[A. J. Nishizawa et al.]
{
  Atsushi J. Nishizawa, Masamune Oguri and Masahiro Takada\\
  Kavli Institute for the Physics and Mathematics of the Universe
  (Kavli IPMU, WPI), The University of Tokyo, Chiba 277-8582, Japan\\
}

\date{\today}

\maketitle

\label{firstpage}
\begin{abstract}
  The baryon acoustic oscillation (BAO) measurement requires a
  sufficiently dense sampling of large-scale structure tracers with
  spectroscopic redshift, which is observationally expensive
  especially at high redshifts $z\simgt 1$.  Here we present an
  alternative route of the BAO analysis that uses the
  cross-correlation of sparse spectroscopic tracers with a much denser
  photometric sample, where the spectroscopic tracers can be quasars
  or bright, rare galaxies that are easier to access
  spectroscopically. We show that measurements of the
  cross-correlation as a function of the transverse comoving
  separation rather than the angular separation avoid a smearing of
  the BAO feature without mixing the different scales at different
  redshifts in the projection, even for a wide redshift slice $\Delta
  z\simeq 1$. The bias, scatter, and catastrophic redshift errors of
  the photometric sample affect only the overall normalization of the
  cross-correlation which can be marginalized over when constraining
  the angular diameter distance.  As a specific example, we forecast
  the expected accuracy of the BAO geometrical test via the
  cross-correlation of the Sloan Digital Sky Survey (SDSS) and 
  Baryon Oscillation Spectroscopic Survey (BOSS) spectroscopic quasar sample
  with a dense photometric galaxy sample that is assumed to have a
  full overlap with the SDSS/BOSS survey region.  We show that this
  cross-correlation BAO analysis allows us to measure the angular
  diameter distances to a fractional accuracy of about 10\% at each
  redshift bin over $1\simlt z\simlt 3$, if the photometric redshift
  errors of the galaxies, $\sigma_z/(1+z)$, are better than $10-20$\%
  level.
\end{abstract}


%
\section{Introduction}
\label{sec:intro}
%

Various cosmological data sets such as the cosmic microwave background
\citep[CMB;][]{Hinshowetal:2012}, the Type Ia supernova observations
\citep{Riessetal:1998,Schmidtetal:1998,Perlmutteretal:1999,Kessleretal:2009,Suzukietal:12}
and the baryon acoustic oscillation (BAO) measurements
\citep{Eisensteinetal:05,Percivaletal:2007,Percivaletal:2010,Beutleretal:2011,Blakeetal:2011,Andersonetal:12}
have shown increasing evidence that the cosmic expansion today is in
the accelerating expansion phase. The cosmic acceleration is the most
tantalizing problem in cosmology.

Among others, the BAO measurement is recognized as one of the most
promising geometrical tests, because it rests on the physics of the
CMB anisotropies in the early universe, which is remarkably well
described by the linearized perturbation theory.  The tight coupling
between baryons and photons prior to the decoupling epoch of $z\simeq
1100$ leaves a characteristic imprint on the pattern of large-scale
structure tracers such as galaxies and quasars -- the so-called BAO
scale.  The BAO scale is now precisely constrained as $\simeq 150~$Mpc
from the CMB observations \citep{Hinshowetal:2012}, which can be used
as a `standard ruler' to infer the cosmological distances from the
observed correlation function of the tracers
\citep{HuHaiman:03,SeoEisenstein:03}.

The BAO measurements mostly utilize a large data from wide-area
redshift surveys of galaxies, such as the
6dF Galaxy Survey (6dFGS)\footnote{http://www.aao.gov.au/6dFGS/}, the Sloan Digital Sky
Survey (SDSS)\footnote{http://www.sdss.org/}, the Baryon Oscillation
Spectroscopic Survey
(BOSS)\footnote{http://www.sdss3.org/surveys/boss.php} and the
WiggleZ survey\footnote{http://wigglez.swin.edu.au/site/}.  With the
success of these surveys, there are several future BAO surveys
targeting higher redshift ranges of $z\simgt 1$, including the 
Hobby-Eberly Telescope Dark Energy Experiment (HETDEX)
survey\footnote{http://hetdex.org}, the Extended Baryon Oscillation
Spectroscopic Survey
(eBOSS)\footnote{http://www.sdss3.org/future/eboss.php}, the
BigBOSS\footnote{http://bigboss.lbl.gov/}, the Subaru Prime Focus
Spectrograph project\footnote{http://sumire.ipmu.jp/en/2652}
\citep{Ellisetal:12}, and the satellite {\it Euclid}
mission\footnote{http://sci.esa.int/science-e/www/object/index.cfm?fobjectid=48983}. However,
extending the BAO measurement to higher redshifts ($z\sim 2-4$) is
observationally expensive, because the target galaxies become
increasingly fainter and spectroscopic surveys of such faint galaxies
having a wide-area coverage
are quite time-consuming\footnote{We note that the BAO feature at
  $z\la 2.3$ has been recently detected from the three-dimensional
  correlation function of the Lyman-$\alpha$ forests that are
  identified in the BOSS quasar spectra
  \citep{Buscaetal:13,Slosaretal:13}.}.

In addition to these spectroscopic BAO analysis, there have been
attempts to measure the BAO feature in the correlation function of
photometric galaxy samples
\citep{Blakeetal:07,Padmanabhanetal:07,Carneroetal:12,Seoetal:12}.  A
wide-area, multi-colour photometric survey is relatively easy to carry
out compared to a spectroscopic survey of similar area coverage. In
fact, there are many planned imaging surveys, including the Subaru
Hyper Suprime-Cam (HSC)
Survey\footnote{http://www.naoj.org/Projects/HSC/index.html}, the Dark
Energy Survey (DES)\footnote{http://www.darkenergysurvey.org}, {\it Euclid}
and the Large Synoptic Survey Telescope (LSST) project\footnote{http://www.lsst.org/lsst/}, for which
the primary science driver is weak lensing based cosmology.  However,
the photometric BAO measurements are challenging for several
reasons. First, the photometric BAO analysis is based on the angular
correlation function of the galaxies, which is by nature
two-dimensional and therefore loses
the clustering information in the line-of-sight direction. 
Secondly, the projection along the line-of-sight 
mixes the different physical scales and smears the BAO feature
in the angular correlation. The projection also reduces the overall
amplitude of the angular correlation function.
Thirdly, the BAO feature inferred from the photometric samples
can be significantly affected by statistical and systematic
(catastrophic) errors of the photometric redshifts (photo-$z$s). For
example, including the photo-$z$ outliers in the analysis can easily
induce a bias in the BAO peaks, which in turn causes a bias in the
inferred distance.

In this paper, we propose to use the cross-correlation between the
spectroscopic and photometric tracers of large-scale structure as an
alternative BAO method. This method is particularly useful when
sampling of the spectroscopic tracers is too sparse to measure the BAO
feature via its auto-correlation analysis.  Since a photometric survey
usually has a much denser sampling, the cross-correlation mitigates
the shot noise contamination to improve clustering measurements.  We
argue that smearing due to the line-of-sight projection can be avoided
by measuring the correlation function as a function of the transverse
comoving separation rather than the angular separation. As a specific
example, we consider the cross-correlation of the SDSS Data Release 7 (DR7)
and BOSS Data Release 9 (DR9; hereafter SDSS/BOSS) spectroscopic sample of quasars with
photometric galaxies to estimate the expected accuracy of the derivable
geometrical test. The SDSS/BOSS quasars are bright and can easily be
observed spectroscopically, but have a too sparse sampling for the
auto-correlation analysis. When making the forecast, we also include
the broad-band shape information of the cross-correlation in addition
to the BAO feature \citep[also see][]{Coorayetal:01}.

This paper is organized as follows. In Section~\ref{sec:corr}, we
describe explicit expressions for the cross-correlation analysis as a
function of the transverse comoving separation as well as its
counterpart in Fourier space, and also derive the covariance matrix.
We show our basic results in Section~\ref{sec:result}. In
Section~\ref{sec:result_fisher}, we show the expected accuracy of the
geometrical test via the use of the cross-correlation of the SDSS/BOSS
spectroscopic quasar sample with a dense photometric galaxy sample.
We summarize our results in Section~\ref{sec:summary}.  Unless
otherwise stated, we employ a concordance $\Lambda$ cold dark matter
($\Lambda$CDM) model \citep{Komatsuetal:10}, with $\Omega_{\rm
  m0}h^2=0.137$ and $\Omega_{\rm b}h^2=0.023$ for the matter and
baryon physical density parameters, $\Omega_\Lambda=0.721$ for the
cosmological constant assuming a flat geometry and $A_{\rm s}=2.43\times
10^{-9}$, $n_{\rm s}=0.96$ and $\alpha_{\rm s}=0$ for the primordial power
spectrum parameters.

%
\section{BAO feature in the projected correlation function}
\label{sec:corr}
%

\subsection{Transverse cross-correlation function and the power spectrum}
\label{ssec:ps}

In this paper, we consider a method that uses the cross-correlation of a
photometric sample with a spectroscopic sample for measuring the BAO
scale. A key idea is to consider the cross-correlation measured as a
function of the transverse comoving separation rather than the
angular separation
\begin{equation}
  w(R)\equiv \frac{1}{\bar{n}_{\rm s}\bar{n}_{\rm p}}
  \left[\ave{n_{\rm s}(\bm{\gamma}_{\rm s}; z_{\rm s})n_{\rm p}(\bm{\gamma}_{\rm p})} -1 \right],
  \label{eq:wR}
\end{equation}
where quantities with subscripts `${\rm s}$' and `${\rm p}$' denote those for
spectroscopic and photometric samples, respectively;
$n_{\rm s}(\bm{\gamma}_{\rm s}; z_{\rm s})$ and $n_{\rm p}(\bm{\gamma}_{\rm p})$ are the projected
number density fields for the spectroscopic and photometric samples in
the directions of $\bm{\gamma}_{\rm s}$ and $\bm{\gamma}_{\rm p}$ on the celestial
sphere, respectively; $z_{\rm s}$ is the redshift of each spectroscopic
object. Thus the density field of the spectroscopic sample is
described as a function of both $z_{\rm s}$ and the angular position. The
transverse radius $R$ is defined in terms of their observed angular
positions of $\bm{\gamma}_{\rm s}$ and $\bm{\gamma}_{\rm p}$ and the redshift
$z_{\rm s}$ as
\begin{equation}
R=d_{\rm A}(z_{\rm s})\cos^{-1}(\bm{\gamma}_{\rm s}\cdot\bm{\gamma}_{\rm p})\simeq
 d_{\rm A}(z_{\rm s})|\bm{\theta}_{\rm s} - \bm{\theta}_{\rm p}|.
\label{eq:R}
\end{equation}
The quantity $d_{\rm A}(z_{\rm s})$ is the comoving angular diameter distance to
each spectroscopic object. Note that this conversion requires to
assume a background cosmological model. The unit vector on the
celestial sphere, $\bm{\gamma}$, is given as $\bm{\gamma}\equiv
(\sin\vartheta\cos\varphi,\sin\vartheta\sin\varphi, \cos\vartheta)$.
In the last equality on the right-hand side (rhs) of equation (\ref{eq:R}), we used the
flat-sky approximation\footnote{In the flat-sky approximation, the
  unit vector is expanded around the North Pole as $\bm{\gamma}\simeq
  (\vartheta\cos\varphi,\vartheta\sin\varphi,1)$, and the
  two-dimensional flat-space vector can be defined as
  $\bm{\theta}\equiv (\vartheta\cos\varphi,\vartheta\sin\varphi)$.}.
Observationally, the cross-correlation is estimated from the average
of all the pairs separated by the same separation $R$ within a given
width, compared to the cross-correlation of the spectroscopic sample
with random catalogues that are constructed based on the same
selection function as in the photometric catalogue.

A notable advantage of the $R$-average over the angle average is that
it can preserve the physical scales inherent in large-scale structure
such as the scale of BAO in which we are
interested.  On the other hand, the $\theta$-average mixes different
scales in large-scale structure, thus smearing the BAO scale in the
observed cross-correlation function. We emphasize that this
$R$-average is useful when a spectroscopic catalogue is available for
the cross-correlation measurement. In contrast, in the case of the
auto-correlation analysis of photometric samples, the conversion from
$\theta$ to $R$ is severely affected by photo-$z$ uncertainties, which
can lead to a smearing and systematic offset of the BAO feature.

We can express the projected cross-correlation function in terms of
the power spectrum as follows. First, 
considering the spectroscopic sample redshift distribution,
the projected cross-correlation can be expressed as
\begin{equation}
  w(R)=\int_0^\infty\!\! \id z_{\rm s}~p_{\rm s}(z_{\rm s}) 
  \left. \tilde{w}(\theta; z_{\rm s})
  \right|_{R=d_{\rm A}(z_{\rm s})\theta},
\end{equation}
where $\tilde{w}(\theta; z_{\rm s})$ is the angular cross-correlation
function of a spectroscopic sample at redshift $z_{\rm s}$ with a
photometric sample, $p_{\rm s}(z_{\rm s})$ is the redshift distribution of the
spectroscopic sample, normalized as 
$\int_0^{\infty} \id z_{\rm s}~p_{\rm s}(z_{\rm s})=1$,
and the average with the notation $\left.{~
  }\right|_{R=d_{\rm A}(z_{\rm s})\theta}$ indicates that the redshift average for
a given $R$ is done by averaging the angular correlation function
$\tilde{w}(\theta; z_{\rm s})$ under the condition $R=d_{\rm A}(z_{\rm s})\theta$
according to the discussion around equations (\ref{eq:wR}) and
(\ref{eq:R}).  $\tilde{w}(\theta; z_{\rm s})$ is defined in terms of the
angular power spectrum as
\begin{equation}
  \tilde{w}(\theta; z_{\rm s})=\frac{1}{4\pi}\sum_l(2l+1)C_{\rm sp}(l;z_{\rm s}) P_l(\cos\theta). 
\end{equation}
Here $P_l(x)$ is the $l$th order Legendre polynomials. For a flat
universe, the angular power spectrum is given in terms of the
three-dimensional power spectrum, in a standard manner
\citep[e.g.,][]{Dodelsonbook}, as
\begin{equation}
  C_{\rm sp}(l;z_{\rm s})
  \equiv
  \frac{2}{\pi}\int\!\id r W_{\rm p}(r)
  \int\!k^2\id k~ P_{\rm sp}(k; z_{\rm s}, z)j_l(kr_{\rm s})j_l(kr),
\end{equation}
where $P_{\rm sp}(k;z_{\rm s},z)$ is the three-dimensional cross-power spectrum
between the spectroscopic objects at redshift $z_{\rm s}$ and photometric
objects at $z$; $r$ is the radial distance given as a function of
redshift for a given cosmology, $r=r(z)$, and $r_{\rm s}=r(z_{\rm s})$; $j_l(x)$
is the $l$th order spherical Bessel function; $W_{\rm p}(r)$ is the
selection function of the photometric sample, normalized as $\int\!\id r
W_{\rm p}(r)=1$ (see below for an example). Note that $r(z)=d_{\rm A}(z)$ for a
flat universe.

Using the flat-sky approximation and the Limber's approximation
\citep{Limber:54}, the projected cross-correlation function can be
simplified as
\begin{equation}
  w(R)\equiv \int\!\frac{k\id k}{2\pi}C_{\rm sp}(k)J_0(kR),
  \label{eq:wRl}
\end{equation}
where $J_0(x)$ is the zeroth-order Bessel function, and the transverse
comoving separation separation $R$ is defined for spectroscopic
redshift $z_{\rm s}$ of each sample used in the average. The projected
cross-power spectrum $C_{\rm sp}(k)$ is given by a simple form:
\begin{equation}
  C_{\rm sp}(k)\equiv \int\!\id r ~ p_{\rm s}(z)\frac{\id z}{\id r}
  W_{\rm p}(r) P_{\rm sp}(k; z). 
  \label{eq:ck_def}
\end{equation}
Note that the power spectrum $C_{\rm sp}(k)$ has a dimension of ${\rm
  Mpc}^2$ so that $k^2C_{\rm sp}(k)$ becomes dimension-less.  We use
$C_{\rm sp}(l)$ for the usual angular power spectrum and $C_{\rm sp}(k)$ for
the Fourier counterpart of the $w(R)$ throughout the paper.  We have
checked that, for the fiducial set-up we study in this paper, the
Limber's approximation is accurate at sub percent level for the BAO
scale.

We assume that we can, based on photo-$z$ technique, select
photometric objects that have similar photo-$z$ to the spectroscopic
redshift. Even in the presence of large photo-$z$ errors, the
cross-correlation method is very powerful in the sense that it can
statistically select photometric objects that are physically
clustering with the spectroscopic sample
\citep{Newman:08,McQuinnWhite:13}.  Including photo-$z$ bias and
outliers in the sample simply dilutes the cross-correlation signals,
but does not change the shape so that the BAO scale is not shifted. If
the spectroscopic sample used in the cross-correlation measurement is
in a narrow range of redshifts, [$z_{\rm s}$ ,$z_{\rm s}+\Delta z_{\rm s}$], the
projected power spectrum reads
\begin{equation}
  C_{\rm sp}(k)
  \simeq 
  \left[
    \int_{r_{\rm s}}^{r_{\rm s}+\Delta r_{\rm s}}\id r\,   W_{\rm p}(r)
  \right] 
  \frac{1}{\Delta r_{\rm s}}P_{\rm sp}(k;z_{\rm s}),
  \label{eq:ck_approx}
\end{equation}
where $r_{\rm s}$ and $r_{\rm s}+\Delta r_{\rm s}$ are the radial
distances to redshifts $z_{\rm s}$ and $z_{\rm s}+\Delta z_{\rm s}$,
respectively, and we have used $\Delta r_{\rm s} p_{\rm s}(z_{\rm
  s})(\id z/\id r)=1$. The factor $\int_{r_{\rm s}}^{r_{\rm s}+\Delta
  r_{\rm s}}~\id r\, W_{\rm p}(r)$ is the fraction of photometric
objects among the whole photometric sample that reside in the
spectroscopic redshift bin $[z_{\rm s},z_{\rm s}+\Delta z_{\rm s}]$
and thus are physically correlated with the spectroscopic
sample. Therefore the factor gives a dilution factor of the
cross-correlation signal due to the photo-$z$ errors.  The factor
$1/\Delta \chi_{\rm s}$ in front of $P_{\rm sp}(k)$ accounts for the
fact that the cross-correlation amplitude is reduced with increasing
the width of spectroscopic redshift bin.  The above equation
explicitly shows that the inclusion of the photo-$z$ outliers does not
change the shape of the cross-correlation, but simply affects the
overall normalization. Also importantly, the projected
cross-correlation, measured against $R$ instead of the angular
separation, can measure the three-dimensional power spectrum $P_{\rm
  sp}(k;z_{\rm s})$ at given spectroscopic redshift and at a
particular $k$. Put another way, the projected cross-correlation does
not mix the power spectrum of different Fourier modes, which is not
the case for the angular power spectrum. It should also be noted that
the projected cross-correlation is not affected by redshift-space
distortion (RSD) due to the peculiar motions of the tracers.  In particular,
the non-linear RSD, the so-called Finger-of-God
effect, is very difficult to accurately model \citep{Hikageetal:12b},
and therefore the cross-correlation may have a practical advantage.

Next, let us consider the case where the spectroscopic sample is in a
given redshift bin with $z_{\rm s}=[z_{\rm s}^{\rm low}, z_{\rm s}^{\rm up}]$,
where $z_{\rm s}^{\rm low}$ and $z_{\rm s}^{\rm up}$ are the lower and upper bound
of the redshift bin, respectively. Here for simplicity we consider a
uniform distribution of the spectroscopic sample within the given
redshift bin
\begin{equation}
  p_{\rm s}(z)=\left\{
    \begin{array}{ll}
      {\displaystyle \frac{1}{\Delta z_{\rm s}}}
      & \mbox{if $z\in[z_{\rm s}^{\rm low},z_{\rm s}^{\rm up}]$,}\\
      0 & \mbox{otherwise},
    \end{array}
  \right.
\end{equation}
where $\Delta z_{\rm s}\equiv z_{\rm s}^{\rm up}-z_{\rm s}^{\rm low}$.
The uniform redshift distribution is not a critical assumption, and
can be easily generalized to a case that the spectroscopic sample has
a non-uniform redshift distribution. For photometric objects used for
the cross-correlation measurement, we would select the objects if the
best-fitting photo-$z$s are in the range of the spectroscopic redshift
bin. Here we employ the simplified assumption that the probability for
photo-$z$s obeys a Gaussian distribution
\begin{equation}
p_{\rm p}(z|z_{\rm p})=\frac{1}{\sqrt{2\pi}\sigma_{\rm z}}\exp\left[
-\frac{(z_{\rm p}-z)^2}{2\sigma_z^2}
\right],
\label{eq:pz_phz}
\end{equation}
where we assumed the photometric objects with the best-fitting
photo-$z$, $z_{\rm p}$, obey a single population, $\sigma_z$ is the
$1\sigma$ photo-$z$ error, and $z$ is its true redshift. The
probability satisfies the normalization $\int^{\infty}_{-\infty} \id
z_{\rm p}~p_{\rm p}(z|z_{\rm p})=\int^{\infty}_{-\infty}\id z~p_{\rm
  p}(z|z_{\rm p})=1$.  Given the distribution, if the photometric
objects whose photo-$z$s are in the range of the spectroscopic
redshift range, $z_{\rm p}\in [z_s^{\rm low},z_s^{\rm up}]$, the probability
distribution for the true redshift is computed as
\begin{eqnarray}
  p_{\rm p}(z |z_{\rm p}\in[z_{\rm s}^{\rm low},z_{\rm s}^{\rm up}])
  &=&
  \int^{z_{\rm s}^{\rm up}}_{z_{\rm s}^{\rm low}} \id z_{\rm p}~
  p_{\rm p}(z|z_{\rm p}) \nonumber \\
  &=&
  \frac{1}{2}
  \left[
    {\rm erf} (x^{\rm up})-{\rm erf}(x^{\rm low})
  \right],
  \label{eq:nzp_prob}
\end{eqnarray}
where ${\rm erf}(x)$ is the error function and $x^{\rm up/low} =
(z_{\rm s}^{\rm up/low}-z)/\sqrt{2} \sigma_z$.  Taking into account the
overall redshift distribution of the photometric sample, we can derive
the redshift distribution of the photometric sample based on the
photo-$z$ selection:
\begin{equation}
  \label{eq:ng_photo}
n_{{\rm p}\in z_{\rm s}}(z)=\frac{1}{2}n_{\rm p}(z)
\left[
{\rm erf}(x^{\rm up})-{\rm erf}(x^{\rm low})
\right],
\end{equation}
where $n_{\rm p}(z)$ is the redshift distribution of the photometric sample
for which we assume $n_{\rm p}(z)=(z^2/2z_0^3)\exp[-z/z_0]$ parametrized by
$z_0$ \citep[see][]{OguriTakada:11}. Throughout the paper we assume
$z_0=0.4$, yielding the mean redshift $\langle z \rangle=1.2$, and
assume the bias parameter of the photometric sample to
$b_{\rm p}=1.5$. Hence, the selection function of the photometric sample in
each $z_{\rm s}$ bin used for the cross-correlation is given as
\begin{eqnarray}
  W_{\rm p}(r)=\frac{n_{{\rm p}\in z_{\rm s}}(z)}{\bar{n}_{{\rm p}\in z_{\rm s}}}\frac{\id z}{\id r},
\end{eqnarray}
where $\bar{n}_{{\rm p}\in z_{\rm s}}$ is the normalization factor, defined as
$\bar{n}_{{\rm p}\in z_{\rm s}}\equiv \int_0^{\infty}\!\id z~ n_{{\rm p}\in z_{\rm s}}(z)$, so as
to satisfy the condition $\int_{0}^{\infty}~\id r~ W_{\rm p}(r)=1$. We will
study how the accuracy of the BAO measurement changes with
quantities such as $\sigma_z$ and $\Delta z_{\rm s}$, as well as the number
densities of spectroscopic and photometric samples.

\begin{figure}
\begin{center}
  \includegraphics[width=\linewidth,clip]{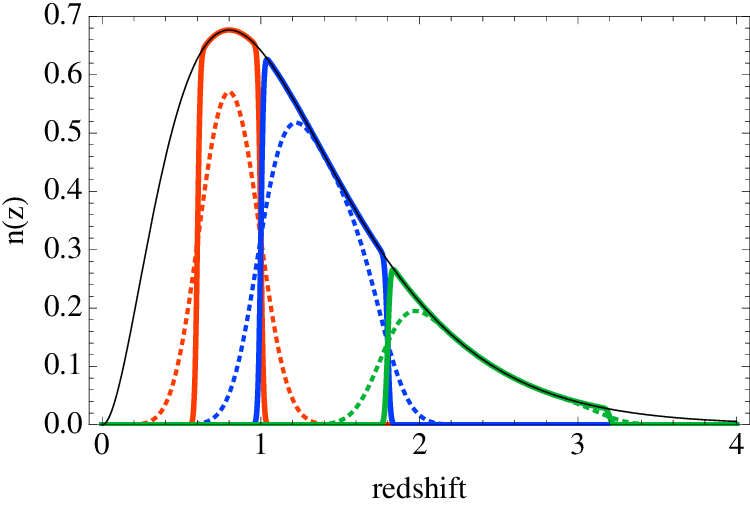}
  \caption{
    The thin solid line shows a toy model of a photometric galaxy
    distribution. Among them, we select galaxies which fall in the
    spectroscopic redshift ($z_{\rm s}$) bin. The thick solid lines show
    the underlying true redshift distribution of the photometric
    samples which are defined by the photo-$z$ bins of 
    $z_{\rm p} \in [0.6,1.0]$, $[1.0,1.8]$ and $[1.8,3.2]$, 
    assuming a photo-$z$ accuracy of $\sigma_z=0.01(1+\bar{z}_s)$ 
    (see Eq.~\ref{eq:pz_phz}).  The dotted lines are similar
    distributions but the photo-$z$ errors are degraded to 
    $\sigma_z=0.1(1+\bar{z}_{\rm s})$. 
    \label{fig:nz}
  }
\end{center}
\end{figure}

In Fig.~\ref{fig:nz}, we show an example of a photometric galaxy
distribution, with different photo-$z$ uncertainties, when the
photometric galaxies are divided in redshift bins, which are chosen
to match with the spectroscopic sample. In this example, we divide the
whole sample into three subsamples as $z_{\rm p} \in [0.6,1.0]$,
$[1.0,1.8]$ and $[1.8,3.2]$. The photo-$z$ errors cause a leakage of the
photometric galaxies from the spectroscopic redshift bin.

\subsection{Covariance matrix}
\label{ssec:cov}

The error covariance matrix quantifies the accuracy of measuring the
projected cross-correlation for a given survey, and is used for the
Fisher matrix analysis presented in Sec.~\ref{sec:result_fisher}.
Since angular scales at different redshifts are scaled to match the
transverse comoving scale for an assumed cosmological model, the
measured projected cross-correlation is two-dimensional, given as a
function of the comoving scales in units of Mpc. Assuming a Gaussian
error for the projected power spectrum, which is a good approximation
at BAO scales \citep{Takahashietal:09}, we can extend the standard
formula for the covariance matrix of angular power spectra
\citep{Knox:95} to obtain the covariance matrix of the projected
cross-correlation function as
\begin{eqnarray}
  &&{\rm Cov}[C_{\rm sp}(k),C_{\rm sp}(k')]
  =
  \frac{\delta^K_{kk'}}{N_{\rm mode}(k)} \nonumber \\
  && \hspace{-0.1cm} \times \left[ 
    C_{\rm sp}(k)^2 +
    \left( 
      C_{\rm ss}(k)+\frac{1}{\bar{n}_{\rm s}}
    \right) 
    \left(
      C_{\rm pp}(k)+\frac{1}{\bar{n}_{{\rm p}\in z_{\rm s}}} 
    \right) 
  \right],
  \label{eq:covariance}
\end{eqnarray}
where $N_{\rm mode}(k)$ is the number of independent Fourier mode 
discriminated by the given survey area defined as
\begin{eqnarray}
  N_{\rm mode}(k)
  &=&
  \frac{2\pi k \Delta k}{\displaystyle 
    \left(\frac{2\pi}{d_{\rm A}(z_{\rm s}^{\rm low})\Theta_{\rm s}
      }\right)^2} \nonumber \\
  &=&
  2k\Delta k~ d_{\rm A}(z_{\rm s}^{\rm low})^2f_{\rm sky},
  \label{eq:nmode}
\end{eqnarray}
with $f_{\rm sky}$ being the sky coverage defined as $f_{\rm
  sky}\equiv \Omega_{\rm s}/4\pi$. The quantities $\bar{n}_{\rm s}$ and $n_{{\rm p}\in
  z_{\rm s}}$ are the projected number densities of the spectroscopic sample
and the photometric sample having photo-$z$'s values within the
spectroscopic redshift bin, respectively. The number densities are in
units of ${\rm Mpc}^{-2}$. In the above equations, we assumed that the
fundamental model of the two-dimensional Fourier decomposition is
defined as the projected scale at the lowest redshift for a given
survey area, $k_{\rm f}\equiv 2\pi/[d_{\rm A}(z_{\rm s}^{\rm low})\Theta_{\rm s}]$.

\begin{figure}
\begin{center}
  \includegraphics[width=\linewidth,clip]{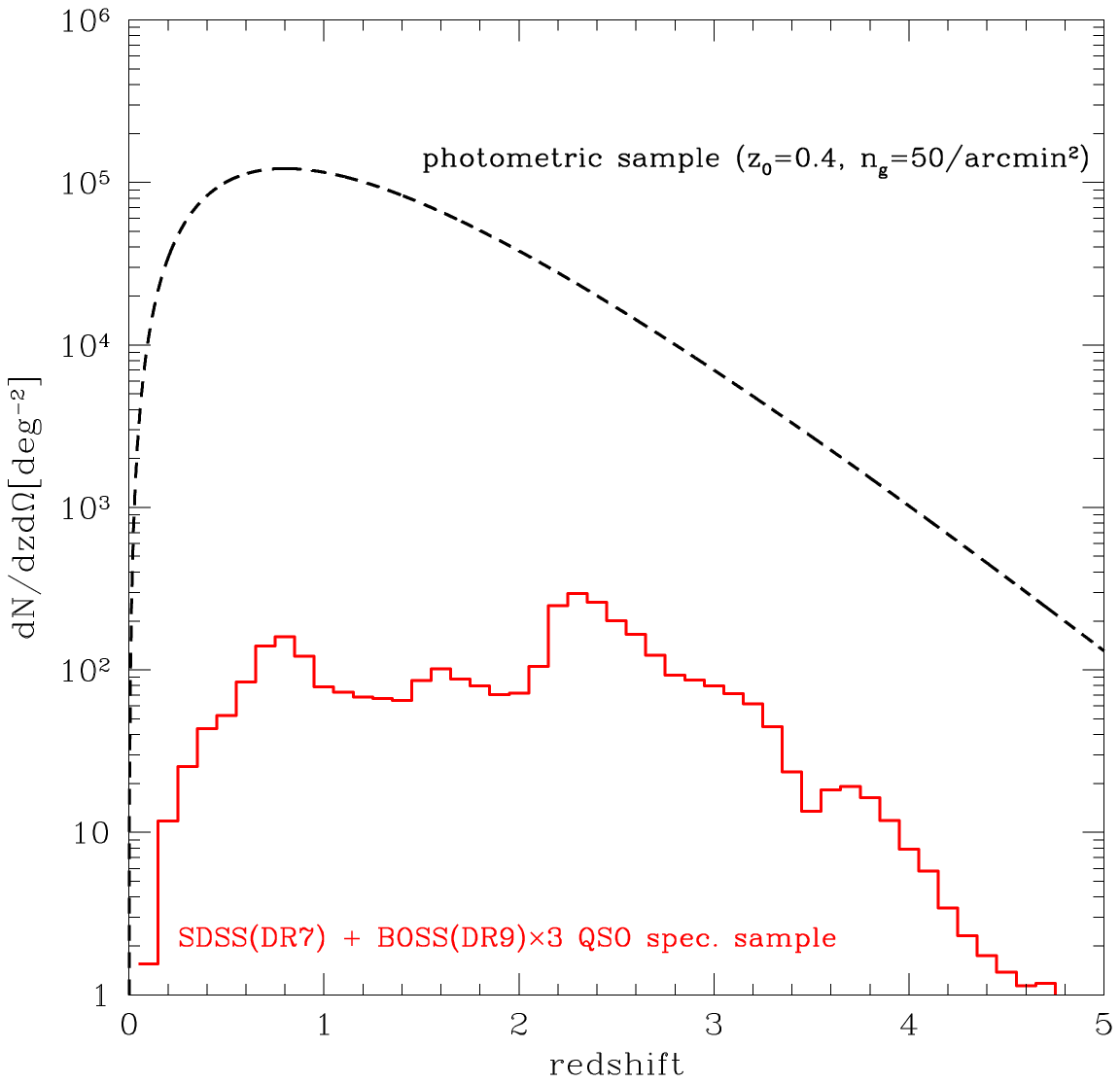}
  \caption{
    The histogram shows the redshift distribution of the spectroscopic
    quasar sample of SDSS DR7 \citep{Schneideretal:10} and BOSS DR9 
    \citep{Parisetal:2012}. The BOSS sample is multiplied by 3 as the 
    DR9 has completed only the 1/3 of the target area. 
    Dashed line shows the assumed 
    photometric galaxy distribution, 
    $n_{\rm p}(z)$ defined below equation (\ref{eq:ng_photo}), where we assumed
    $\bar{n}_{\rm p}=50$~arcmin$^{-2}$ for the total mean number density and
    $\langle z\rangle=1.2$ for the mean redshift.
    \label{fig:dndz_qso}
  }
\end{center}
\end{figure}

\begin{figure*}
\begin{center}
  \includegraphics[width=\linewidth]{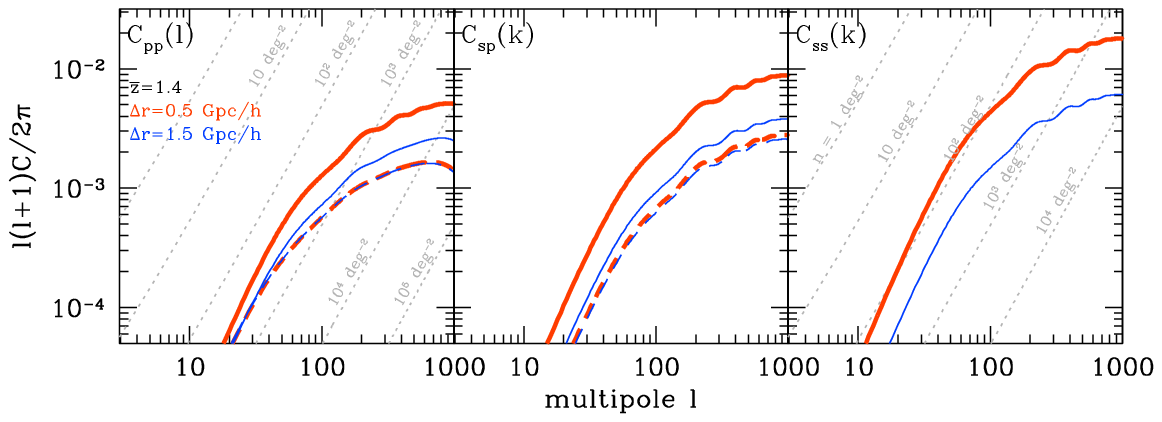}
  \caption[angular power spectra]{
    Comparison of the projected and angular power spectrum at the mean 
    redshift $\bar{z}=1.4$. {\em Left}: the angular autopower
    spectrum of photometric samples in the photo-$z$ bins
    $z_{\rm p}=[1.4-\Delta z/2,1.4+\Delta z/2]$. The thick and thin curves
    show the power spectrum for the bin width of $\Delta z_{\rm p}\simeq
    0.365$ and 1.05, which corresponds to the width of the comoving
    radial distance, $\Delta r=0.5$ and $1.5$~Gpc$/h$, respectively.
    The solid and dashed curves are the spectra assuming the
    photo-$z$ accuracies of $\sigma_z/(1+z)=0.05$ or 0.3, receptively. 
    Each thin dotted lines show the shot noise level for the
    photometric samples, which typically have the projected number
    density more than $10^4~{\rm deg}^{-2}$ for an imaging survey we
    are interested in. {\em Middle}: similar to the left-hand panel, but
    for the cross-power spectrum between the spectroscopic and
    photometric samples, as a function of the transverse comoving
    separation (equation \ref{eq:ck_def}), where the transverse mode $k$ is
    rescaled to the multipole via the distance to the spectroscopic
    sample by $l=kr(z=1.4)$ for an illustrative purpose. 
    The solid and dashed
    curves are for the photo-$z$ accuracies of the photometric
    galaxies, as in the left-hand panel. The cross-correlation preserves
    the BAO wiggles compared to the left-hand panel. {\em Right}: the
    projected auto-power spectrum for the spectroscopic samples.
    The figure shows that, for a spectroscopic survey with a small
    number density  $\bar{n}_{\rm s}<10^{2}~{\rm deg}^{-2}$, the BAO wiggles
    in the auto-spectrum are difficult to measure due to the
    significant shot noise.
    \label{fig:cl}
  }
\end{center}
\end{figure*}
\begin{figure}
  \begin{center}
    \includegraphics[width=0.8\linewidth]{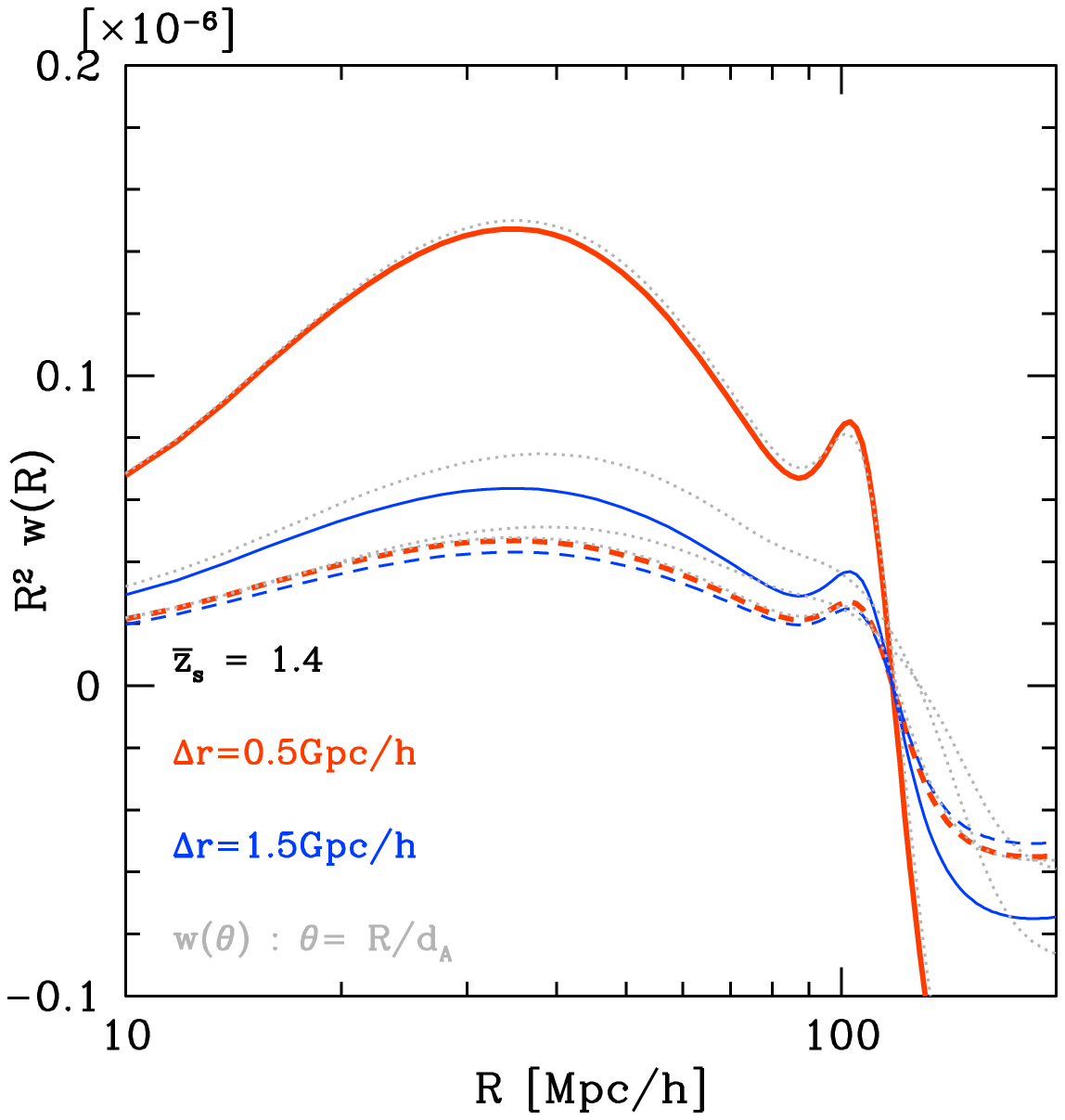}
    \caption{ Similar to Fig.~\ref{fig:cl}, but for the
        cross-correlation function in configuration space, $w(R)$.  As
        in Fig.~\ref{fig:cl}, the top and dashed curves are the
        cross-correlation function assuming the photo-$z$ errors of
        $\sigma_z/(1+z)=0.05$ and 0.3, respectively. The two curves
        differ for the redshift widths; the wider bin width changes
        only the amplitude of $w(R)$, but preserve the overall shape
        and BAO feature. For comparison, the dotted lines show the
        angular cross-correlation $w(\theta=R/d_{\rm A})$ for the same width
        of the (photometric) redshift bin; the BAO peak is
        significantly smeared.
      \label{fig:wR}
    }
  \end{center}
\end{figure}

The expression for the covariance matrix (equation \ref{eq:covariance}) can
be used to understand why the cross-correlation can be useful for the
BAO analysis when the spectroscopic catalogue has too sparse sampling
of the targets, i.e., $C_{\rm ss}\ll 1/\bar{n}_{\rm s}$.  We assume that the
photometric sample has a high number density in the spectroscopic
redshift bin, i.e., $C_{\rm pp}\gg 1/\bar{n}_{{\rm p}\in z_{\rm s}}$, at the BAO
scale, and the cross-correlation coefficient between the spectroscopic
and photometric samples, $r\equiv C_{\rm sp}/\sqrt{C_{\rm ss}C_{\rm pp}}$, is of
order unity.  In this case, the covariance for the cross-correlation
becomes ${\rm Cov}[C_{\rm sp},C_{\rm sp}]\propto C_{\rm pp}/\bar{n}_{\rm s}$, and thus
the signal-to-noise ratio is $({\rm S/N})^2_{\rm sp}=C_{\rm sp}^2{\rm
  Cov}^{-1}\propto r^2C_{\rm ss}\bar{n}_{\rm s}$.  This should be compared with
the case of the autopower spectra of the spectroscopic sample, $({\rm
  S/N})^2_{\rm ss} \propto (C_{\rm ss}\bar{n}_{\rm s})^2$, yielding the S/N ratio
$({\rm S/N})_{\rm sp}^2/({\rm S/N})^2_{\rm ss} \simeq r^2/(C_{\rm ss}\bar{n}_{\rm s})
\gg 1$.  Thus the cross-correlation can give a higher ${\rm S/N}$
ratio for a measurement of the projected power spectrum at the BAO
scales.  In practice, however, the spectroscopic sample allows a
measurement of the three-dimensional power spectrum, which contains
more Fourier modes than in the projected power spectrum. In the next
section, we present a more quantitative comparison between the methods
using the cross-correlation and the three-dimensional
auto-correlation.

Fig.~\ref{fig:dndz_qso} shows the photometric and spectroscopic
samples for which we think the cross-correlation method discussed in
this paper is useful, {\em if} the two survey regions are
overlapped. The spectroscopic quasar catalogues of the SDSS/BOSS
surveys have a wide coverage of redshifts up to $z_s\simeq 4$, but have
a much lower number density than in the photometric galaxies available
from the upcoming imaging surveys such as the Subaru HSC Survey or
{\it Euclid}.  The redshift distribution of the photometric sample shown in
Fig.~\ref{fig:dndz_qso} is deeper than what is usually assumed for the
HSC survey or {\it Euclid}, but we note that for the cross-correlation
analysis we can use fainter galaxies than galaxies used for the weak
lensing analysis. We also note that our results are not very sensitive
to the choice of the number density distribution of the photometric
sample.

%
\section{Results}
\label{sec:result}
%
\subsection{Projected power spectrum}
\label{ssec:result_power}

In Fig.~\ref{fig:cl} we compare the auto- and cross-power spectra for
spectroscopic and photometric samples at mean redshift
$\bar{z}=1.4$. Here the cross-power spectra are computed as a function
of the transverse comoving separation as described in the previous
section. Here, we consider the redshift bin around $z=1.4$,
$z=[z-\Delta z/2,z+\Delta z/2]$ with widths $\Delta z=0.365$ and 1.05,
which correspond to the radial distance widths of $\Delta r=0.5$ and
$1.5$~{\rm Mpc}$/h$, respectively.  To model the photo-$z$ errors, we
use the parametrization given in \citet{Maetal:06} as
\begin{equation}
  \sigma_z 
  = 
  \lambda_z(1+\bar{z}_{\rm s}),
  \label{eq:photoz_accuracy}
\end{equation}
and consider the two cases of $\lambda_z=0.05$ and $0.3$. The BAO
feature is smeared in the angular auto-power spectra of photometric
samples, while the BAO feature persists in the projected auto- or
cross-power spectra using the spectroscopic sample (see also below).
Comparing the solid and dashed curves shows that the larger photo-$z$
errors cause a more significant dilution of the power spectrum
amplitudes, thereby smearing the BAO oscillatory feature.  The figure
also shows the shot noise levels. For ongoing or upcoming imaging
surveys, we typically have more than $10^4 {\rm deg}^{-2}$ galaxies
(see Fig.~\ref{fig:dndz_qso}), and thus the power spectrum measurement
has a sufficient ${\rm S/N}$ ratio. However, the photo-$z$ errors
dilute the spectrum amplitude and smear the BAO feature, suggesting
that it would be difficult to use the angular auto-power spectrum of
the photometric galaxies for an unbiased BAO geometrical test, as we
will discuss below. While the projected cross-power spectra shown in
the figure also have a more diluted amplitude as photo-$z$
uncertainties increase (see equation \ref{eq:ck_approx}), we can still use
the unsmeared BAO feature for estimating the angular diameter
distance.

To be comprehensive, we also show the expected cross-correlation
function in configuration space, $w(R)$, instead of the power spectrum
in Fig.~\ref{fig:wR}.  As in the power spectrum, the overall shape and
the BAO feature are preserved in the $R$-average case, whereas the BAO
peak is significant smeared in the angle-average.

\subsection{Forecast for the cross-correlation BAO measurement}
\label{ssec:result_detection}

\begin{figure}
\begin{center}
  \includegraphics[width=\linewidth]{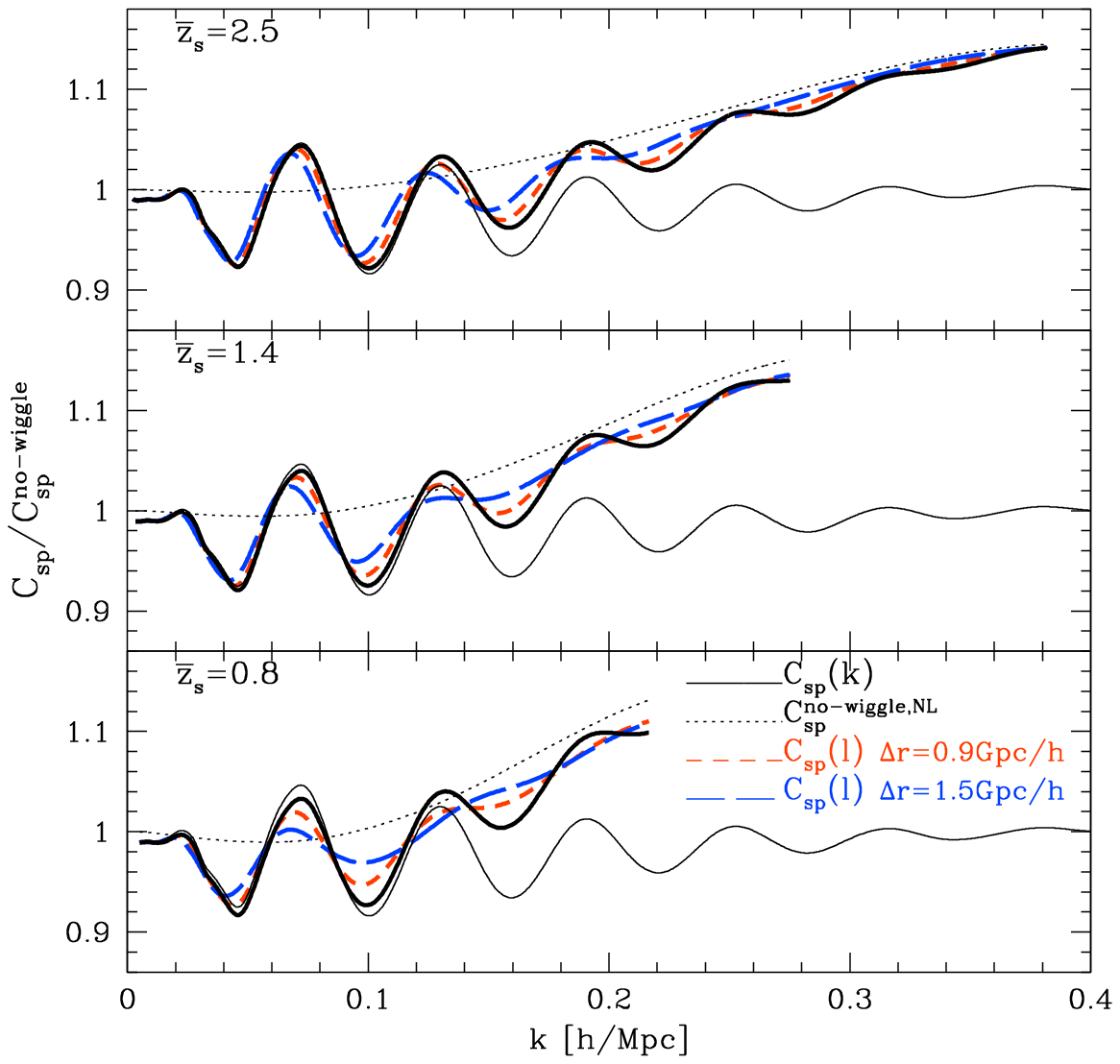}
  \caption[nowiggle normalized power]{
 The projected power spectrum divided by the no-wiggle linear power
 spectrum in order to highlight the BAO feature, where 
 we used \citet{EisensteinHu:98} to compute the no-wiggle spectrum.
 We consider $\bar{z}_{\rm s}=2.5$ ({\it top panel}), $1.4$ ({\it middle})
 and $0.8$ ({\it bottom}), respectively, for the mean redshift of the
 projection. The thick curves are the spectra computed 
 when the non-linearity of the matter power spectrum is considered (see text 
 for the details), while thin curves show the cross-power spectra in
 linear theory. The spectra are plotted up to $k_{\rm max}$, where the
 non-linear power spectra are expected to be accurate to within 1\% at
 each mean redshift compared to simulations. For comparison, we also
 show the angular cross-power spectra projected over the redshift
 slice of $\Delta r=0.9~{\rm Gpc}/h$ ({\it short-dashed}) and 
 $\Delta r=1.5~{\rm Gpc}/h$ ({\it long-dashed}) around each mean
 redshift, respectively, which are plotted against the wavenumber
 using the conversion $kr(\bar{z})=l$. The dotted curves are the
 non-linear power spectrum using the no-wiggle linear power spectrum
 for the input spectrum.
 \label{fig:clnormalized} }
\end{center}
\end{figure}

In this section, we study forecasts for the use of the projected power
spectrum for measuring the BAO feature. In
Fig.~\ref{fig:clnormalized}, we show the projected cross-power
spectrum as a function of the transverse wavenumber, divided by the
no-wiggle power spectrum (with the BAO feature being smoothed out), in
order to highlight the BAO feature. Note that we used the transfer
function in \citet{EisensteinHu:98} to compute the no-wiggle spectrum
for the same cosmological model. Although we assume a linear galaxy
bias multiplicative-factor for both the spectroscopic (with bias
parameter values based on those measured for SDSS quasars;
\citealt{Rossetal:2009}) and photometric samples ($b_{\rm p}=1.5$), we
include the effect of nonlinear clustering on the matter power
spectrum, using the publicly available code, RegPT
\citep{Taruyaetal:2012}, that includes up to the two-loop order
contributions based on the refined perturbation theory. We show the
cross-power spectra up to a certain maximum wavenumber, $k_{\rm max}$,
which is determined so that the non-linear matter power spectrum at the
mean redshift is expected to be accurate to within a 1 per cent level
accuracy in the amplitude compared to the simulation
\citep{Taruyaetal:09,Taruyaetal:2012}.  The figure clearly shows that
the projected cross-power spectrum preserves the BAO feature, even for
a wide redshift bin. On the other hand, the BAO
feature is smeared in the angular correlation. We also notice that,
for the higher redshift slice, the BAO feature remains up to the
greater wavenumber due to the less evolving nonlinearities.

We estimate forecasts for detecting the BAO feature in the projected
cross-spectrum by using the $\chi^2$ difference between the power
spectra with and without the BAO feature:
\begin{equation}
  \Delta \chi^2
  \equiv 
  \sum_i
  \frac{
    \left[
      C_{\rm sp}(k_i)-C_{\rm sp}^{\rm nw}(k_i)
    \right]^2
  }
  {
    {\rm Cov}[C_{\rm sp}(k_i),C_{\rm sp}(k_i)]
  },
  \label{eq:dchi2}
\end{equation}
where $C_{\rm sp}$ and $C_{\rm sp}^{\rm nw}$ are the cross-power spectra with
and without the BAO feature, and the summation is up to the maximum
wavenumber determined as in Fig.~\ref{fig:clnormalized}. Note that
$\Delta\chi^2$ does not include the broad-band shape information of
the cross-power spectrum, and only quantifies the significance of detecting
the BAO feature in the cross-power spectrum, assuming that the
spectrum with the BAO wiggles is the underlying true spectrum (see the
next section for a more quantitative forecast of the BAO analysis).
The denominator is the covariance matrix (equation ~\ref{eq:covariance}) for
which we assumed the Gaussian error. To compute $\Delta\chi^2$, we
assume that the spectroscopic sample has a projected number density of
$\bar{n}_{\rm s}=(20\Delta z)$~\sqdeg in the redshift bin, where $\Delta z$
denotes the redshift bin width. For the photometric sample, we employ
$\bar{n}_{\rm p}^{\rm tot}=1.8 \times 10^5~{\rm deg}^{-2}$
($50~$arcmin$^{-2}$) for the {\it total} number density. These numbers
roughly resemble the SDSS/BOSS quasar spectroscopic sample and the
Subaru HSC Survey or {\it Euclid} imaging surveys, respectively.  We here
assume a full-sky coverage ($f_{\rm sky}=1$) for both the
spectroscopic and photometric catalogues. We note that the chi-square
difference scales as $\Delta\chi^2\propto f_{\rm sky}$ (see
equations (\ref{eq:covariance}), (\ref{eq:nmode}) and (\ref{eq:dchi2})).
We then assume that we can select the photometric objects, which have
photo-$z$'s in the spectroscopic redshift bin, and take into account
the redshift distribution of photometric objects as well as the effect
of photo-$z$ errors using the method in Sec.~\ref{ssec:ps}.

\begin{figure*}
\begin{center}
  \includegraphics[width=0.75\linewidth]{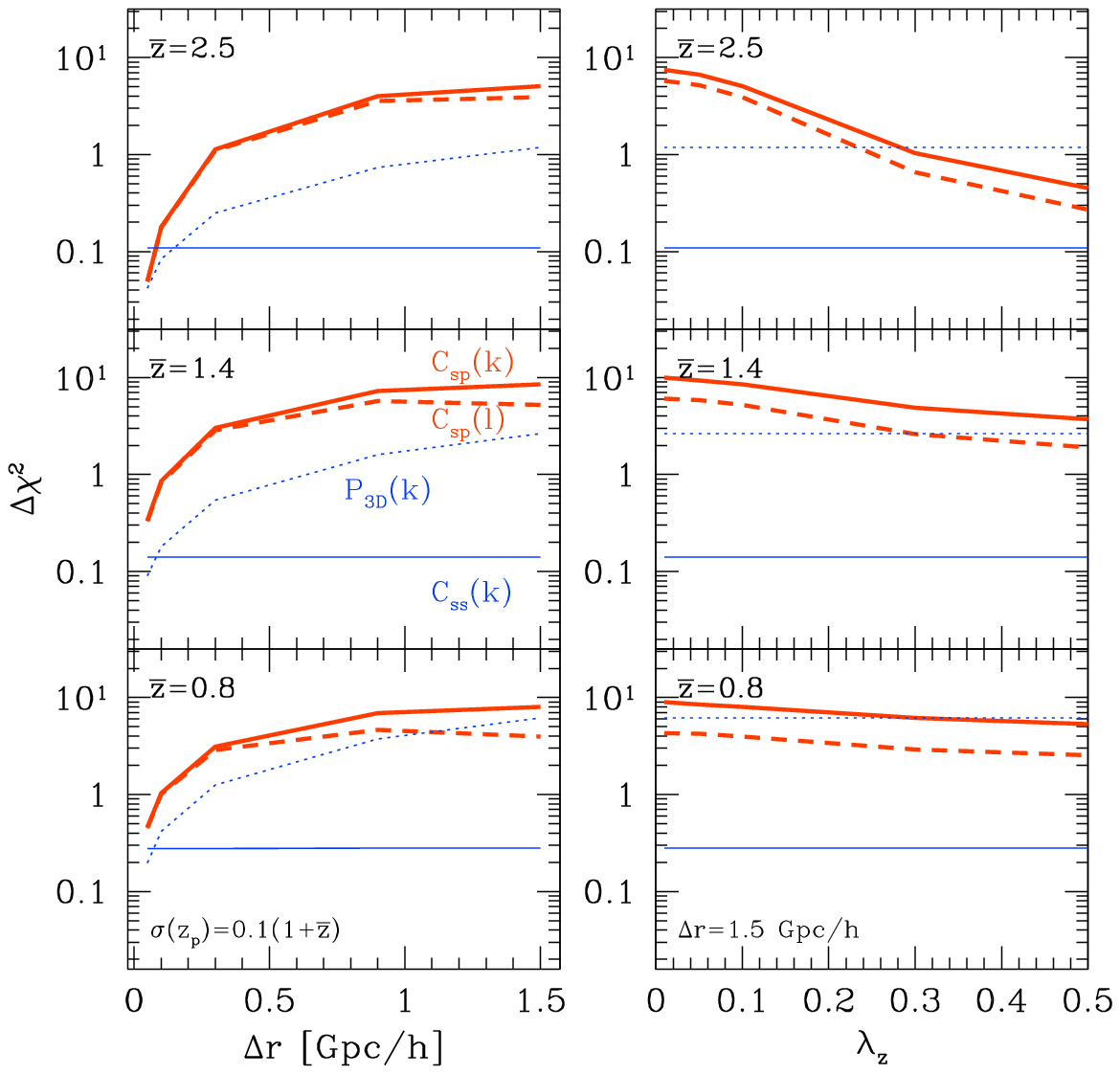}
  \caption[delta chi square]{
    The expected significance of the BAO detection, $\Delta \chi^2$
    (equation \ref{eq:dchi2}), for the cross-correlation analysis with
    different combinations of spectroscopic and photometric
    samples. We estimate the significance by comparing the cross-power
    spectra with and without the BAO wiggles, as in
    Fig.~\ref{fig:clnormalized}, but do not include the broad-band
    shape of the power spectrum. In each panel, the thick solid curves
    show the $\Delta\chi^2$ values for the projected cross-power
    spectrum  ($C_{\rm sp}(k)$), the thick dashed curves are for the
    angular cross-power spectrum ($C_{\rm sp}(l)$), and the thin
    horizontal line is for the projected auto-correlation of the
    spectroscopic sample ($C_{\rm ss}(k)$). For comparison, we also
    show the result when the BAO feature is extracted from the
    three-dimensional power spectrum analysis ($P_{\rm 3D}(k)$), which
    is estimated using equation (\ref{eq:dchi2_3d}). The number density of
    the spectroscopic sample is fixed to $(20\Delta z)$~\sqdeg, and
    the {\it total} number density of the photometric sample is
    assumed to $\bar{n}_{\rm p}^{\rm tot} = 50~{\rm arcmin}^{-2}$.  Results
    are shown for three mean redshift, $\bar{z}_{\rm s}=2.5$ ({\it top
      panels}), $1.4$ ({\it middle}) and $0.8$ ({\it bottom}).  {\it
      Left:} $\Delta \chi^2$ is calculated under the conditions that
    the photo-$z$ accuracy is fixed to $\lambda_z=0.1$ and the
    redshift bin width is varied from $\Delta r=0.1$ to $1.5~{\rm
      Gpc}/h$.  {\it Right:} $\Delta \chi^2$ is calculated with a
    fixed redshift bin width $\Delta r=1.5~{\rm Gpc}/h$, but with
    varying photo-$z$ accuracies. 
    \label{fig:delta_chisq} }
\end{center}
\end{figure*}

Fig.~\ref{fig:delta_chisq} shows the $\Delta \chi^2$ values for the
cross-power spectrum assuming various combinations of the survey
parameters. If the two surveys have a sufficiently wide area coverage
for their overlapping region, the projected cross-power spectrum
allows a detection of the BAO feature. We compare the results with a
BAO analysis for the spectroscopic sample alone. Similarly to
equation (\ref{eq:dchi2}), we can define the differential $\chi^2$ to
quantify the sensitivity of the three-dimensional power spectrum to
the BAO feature
\begin{equation}
  \Delta \chi^2_{\rm 3D}
  \equiv 
  \frac{1}{V_{\rm survey}}\int \frac{2\pi k^2 \id k}{(2\pi)^3} 
  \left[\frac{ P_{s}(k)-P_{s}^{\rm nw}(k) }
  {P_{s}(k)+\bar{n}_s^{-1}} \right]^2,
\label{eq:dchi2_3d}
\end{equation}
where $P_{\rm s}(k)$ is the three-dimensional power spectrum of the
spectroscopic sample and $V_{\rm survey}$ is the survey volume.  We
ignore the RSD for simplicity \citep{Kaiser:87}.
The figure shows that, if the photo-$z$ accuracies of
$\sigma(z_{\rm p})/(1+z)$ are better than 10-20 per cent, the 
cross-correlation can achieve a more significant detection of the 
BAO feature than in the three-dimensional power spectrum.

%
\section{Geometrical test with the cross-correlation method}
\label{sec:result_fisher}
%

\begin{table*}
  \begin{center}
    \begin{tabular}{ccccccclcc}
      \hline \hline
     $\bar{z_s}$            & 
     $\Delta z_s$           & 
     $b_s$                  & 
     $\beta(\bar{z_s})$     & 
     $\bar{n}_s$ (deg$^{-2}$)       &
     $k_{\rm max}$ ($h$/Mpc)        &
     Area (deg$^2$)                & 
     $\lambda_z$                   & 
     $\bar{n}_p$ ($10^4$deg$^{-2}$) &
     $\sigma_{D_{\rm A}}/D_{\rm A}$ \\
     \hline
     0.7 & 0.2 & 1.52 & 0.352 & 3   & 0.21 & 10,000 & 0.01 &  2.4 & 0.076 \\
         &     &      &       &     &      &        & 0.1  &  2.2 & 0.095 \\
         &     &      &       &     &      &        & 0.3  &  1.7 & 0.132 \\
     &&&&&&&\multicolumn{2}{l}{Spec auto correlation}             & 0.191 \\
     \hline                                                  
     0.9 & 0.2 & 1.70 & 0.333 & 3   & 0.23 & 10,000 & 0.01 &  2.4 & 0.095 \\
         &     &      &       &     &      &        & 0.1  &  2.3 & 0.095 \\
         &     &      &       &     &      &        & 0.3  &  1.7 & 0.137 \\
     &&&&&&&\multicolumn{2}{l}{Spec auto correlation}             & 0.237 \\
     \hline                                                  
     1.2 & 0.4 & 2.01 & 0.299 & 3.5 & 0.25 & 10,000 & 0.01 &  4.0 & 0.084 \\
         &     &      &       &     &      &        & 0.1  &  3.9 & 0.098 \\
         &     &      &       &     &      &        & 0.3  &  3.0 & 0.141 \\
     &&&&&&&\multicolumn{2}{l}{Spec auto correlation}             & 0.369 \\
     \hline                                                  
     1.6 & 0.4 & 2.49 & 0.252 & 3.5 & 0.29 & 10,000 & 0.01 &  2.7 & 0.080 \\
         &     &      &       &     &      &        & 0.1  &  2.7 & 0.103 \\
         &     &      &       &     &      &        & 0.3  &  2.4 & 0.188 \\
     &&&&&&&\multicolumn{2}{l}{Spec auto correlation}             & 0.475 \\
     \hline                                                  
     2.2 & 0.8 & 3.36 & 0.193 & 10  & 0.35 & 10,000 & 0.01 &  2.3 & 0.068 \\
         &     &      &       &     &      &        & 0.1  &  2.6 & 0.084 \\
         &     &      &       &     &      &        & 0.3  &  3.2 & 0.188 \\
     &&&&&&&\multicolumn{2}{l}{Spec auto correlation}             & 0.302 \\
     \hline                                                   
     2.9 & 0.6 & 4.60 & 0.144 & 10  & 0.42 & 10,000 & 0.01 &  0.5 & 0.075 \\
         &     &      &       &     &      &        & 0.1  &  0.6 & 0.133 \\
         &     &      &       &     &      &        & 0.3  &  1.4 & 0.536 \\
     &&&&&&&\multicolumn{2}{l}{Spec auto correlation}             & 0.306 \\
     \hline \hline
\end{tabular}
    \caption[survey summary]{A summary of survey parameters we consider
   for the forecast, and the expected fractional errors of determining
   the angular diameter distance, $\sigma(D_A)/D_A$, including
   marginalization over the other parameters. Here we consider the
   SDSS/BOSS spectroscopic quasar catalogue for the spectroscopic 
   sample, and the Subaru HSC- or {\it Euclid}-type galaxy sample for the
   photometric sample.
   $\bar{z_{\rm s}}$ and
   $\Delta z_{\rm s}$ are the mean redshift and the redshift width for each
   redshift bin of the spectroscopic sample taken in the hypothetical
   cross-correlation analysis. $b_{\rm s}$, $\beta$ and $\bar{n}_{\rm s}$ are the
   linear bias, the linear RSD and the number
   density in each redshift bin (see text for the details). 
   $k_{\rm max}$ is the maximum wavelength used for the Fisher matrix
   analysis. For each redshift bin, we cross-correlate the
   spectroscopic sample with the photometric galaxies based on their
   photo-$z$s assuming the photo-$z$ errors of $\lambda_z=0.01$,
   $0.1$ and $0.3$, respectively (see equation \ref{eq:photoz_accuracy}).
   $n_{\rm p}$ is the number density of the photometric galaxies in each
   redshift bin (see equation \ref{eq:ng_photo}). The last column 
   ($\sigma_{D_{\rm A}}/D_{\rm A}$) shows the expected error on the angular
   diameter distance measurement in each bin. For comparison, we also
   show the expected error when using the three-dimensional auto-power
   spectrum of the spectroscopic sample (``spec auto-correlation''). 
      \label{tab:forecasts}
    }
  \end{center}
\end{table*}

In this section, we present more quantitative estimates on the power
of the cross-correlation method for determining the angular diameter
distance. For this forecast, in contrast to the preceding section, we
include the broad-band shape information of the cross-power spectrum,
extending the method in \citet{SeoEisenstein:03} to a two-dimensional
cross-correlation analysis. As a specific example, here we consider
the cross-correlation BAO analysis assuming the SDSS/BOSS
spectroscopic quasar catalogues
\citep{Schneideretal:10,Parisetal:2012} as the spectroscopic sample (as
shown in Fig.~\ref{fig:dndz_qso}) and a mock photometric sample which 
has full overlap with the spectroscopic sample, as the photometric sample.
We assume the total area of
10000~deg$^2$ for the overlapping area.  We consider six redshift
bins with the mean redshifts ranging from 0.7 to 2.9. The projected
number density in each bin is estimated using the redshift
distribution in Fig.~\ref{fig:dndz_qso}. We use the bias parameters of
the quasars in each redshift bin based on the measurement by
\citet{Rossetal:2009}. For the photometric sample, we again assume the
total number density of $\bar{n}_{\rm p}^{\rm tot} = 50~{\rm arcmin}^{-2}$,
and compute the number density in each redshift bin taking into
account the photo-$z$ error (see Sec.~\ref{ssec:ps}).
Table~\ref{tab:forecasts} summarizes the set of the survey parameters.

The cross-correlation is measured as a function of the transverse
separation between the pairs of the spectroscopic and photometric
objects. The transverse separation, the separation distance between
each pair perpendicular to the line-of-sight direction, can be
inferred from the observed angular separation on the sky, $R\propto
\Delta \theta$ (see equation \ref{eq:R}). For this conversion, we need to
assume a reference cosmological model to relate the observable $\Delta
\theta$ to the quantity $R$. Thus the transverse wavenumber is given
as
\begin{equation}
  k_{\bot,{\rm ref}}
  =
  \frac{D_A(z)}{D_{A,{\rm ref}}(z)} 
  k_{\bot}.
  \label{eq:k_perp}
\end{equation}
The quantities with ``ref'' are the quantities from the observable
assuming a ``reference'' cosmological model, and the quantities
without the subscript denote the underlying true quantities. Since the
reference cosmological model assumed generally differs from the
underlying true cosmology, it causes an apparent shift in the
cross-power spectrum.  Thus the observed cross-power spectrum is given
as
\begin{equation}
  C^{\rm obs}_{sp}(k_{\bot ,{\rm ref}}; z)
  =
  \frac{D_{{\rm A},{\rm ref}}(z)^2}{D_{\rm A}(z)^2}
  C_{\rm sp}(k; z) + {\mathcal P}_{\rm s}(z),
  \label{eq:ps_cross}
\end{equation}
where ${\mathcal P}_{\rm s}(z)$ is the residual shot noise\footnote{Suppose
  that the spectroscopic and photometric samples reside in their host
  haloes, which have the number densities of $n_{h1}$ and $n_{h2}$,
  and assume that some fractions of the two samples have the common
  host haloes which have the number density of $n_{c}$. In this case,
  the shot noise for the cross-power spectrum is found to be
  proportional to $n_c/(n_{h1}n_{h2})$.}.  Since we consider a wide
redshift bin for the projection of the spectroscopic sample, we can
safely ignore the RSD effect.

To make a parameter forecast, we include the following set of
parameters: 
\begin{equation}
  \bvec{\theta}=
  \{\Omega_{\rm m}, \Omega_{\rm m}h^2, \Omega_{\rm b}h^2, A_{\rm s}, n_{\rm s},
  \alpha_{\rm s}, D_{\rm A}(z_i), {\mathcal A}(z_i), {\mathcal P}_{\rm s}(z_i)\}, 
\end{equation}
where $\Omega_{\rm m}h^2$ and $\Omega_{\rm b}h^2$ are the matter and baryon
density parameters today, $A_{\rm s}$, is the amplitude of the primordial
curvature perturbation at $k_{\rm piv}=0.005~{\rm Mpc}$ and $n_{\rm s}$ and
$\alpha_{\rm s}$ are the tilt and running of the primordial power spectrum
\citep{Komatsuetal:10}. The parameter $D_{\rm A}(z_i)$ is the angular
diameter distance to the $i$th redshift bin which is treated as an
independent parameter from other cosmological parameters \citep[see
e.g.,][for details]{SeoEisenstein:03,Ellisetal:12}.  We also include
the normalization parameter ${\mathcal A}(z_i)$ which models an
uncertainty in the normalization of the cross-power spectrum in each
redshift bin due to unknown bias uncertainties for both the
photometric ($b_{\rm p}$) and spectroscopic ($b_{\rm s}$) samples. As discussed
above, the marginalization over ${\mathcal A}(z_{\rm i})$ also takes account
of photo-$z$ uncertainties. ${\mathcal P}_{\rm s}(z_{\rm i})$ is a nuisance
parameter to model the residual shot noise parameter. In addition to
these parameters, we include the optical depth and angular diameter
distance to the last scattering surface, $\tau$ and $D_{{\rm A}, {\rm
    CMB}}$, respectively, to the Fisher matrix to describe the CMB
prior.

The full Fisher matrix can be expressed by a simple sum of two Fisher
matrices, $\bvec{F}=\bvec{F}^{\rm CMB}+\bvec{F}^{\rm CC}$, where
$\bvec{F}^{\rm CC}$ denotes the Fisher matrix from the
cross-correlation measurement:
\begin{equation}
  F_{\alpha\beta}^{\rm CC}
  =
  \sum_{i}^{N_{\rm bin}}
  \sum_{k_n=k_f}^{k_{\rm max}}
  \frac{\partial C_{\rm sp}(k_n,z_i)}{\partial \theta_\alpha}
  {\rm Cov}^{-1}
  \frac{\partial C_{\rm sp}(k_n,z_i)}{\partial \theta_\beta},
  \label{eq:fisher_cc}
\end{equation}
where Cov is the covariance matrix given by equation (\ref{eq:covariance}).
The maximum wavenumber $k_{\rm max}$ is set to the maximum scale up to
which the non-linear matter power spectrum at the mean redshift is
expected to be accurate to within 1\% level as in
Fig.~\ref{fig:clnormalized}.  As the large-scale structure has less
non-linearity at higher redshifts, we can theoretically model the
cross-power spectrum more accurately up to the larger wavenumber,
enabling tighter constraints on the angular diameter distances.

For comparison, we also show a forecast for using the
three-dimensional power spectrum of the spectroscopic sample to
estimate the cosmological distances, $H(z_i)$ and $D_{\rm A}(z_i)$. We
follow the methods in \citet{SeoEisenstein:07} \citep[also
see][]{Ellisetal:12}. We model the RSD
\citep{Kaiser:87} and its non-linear effects
\citep{Eisensteinetal:07a} for the three-dimensional power spectrum
with additional parameters; $\beta_i = \id\ln D(z_i)/\id\ln a / b_{\rm s}$ and
$H(z_i)$.  The fiducial value of $\beta$ is listed in
Table~\ref{tab:forecasts}. However the results are not sensitive to
the details, because the power spectrum information at relevant
wavenumber bins is limited by the shot noise contamination for the
sparse spectroscopic sample we are interested in.

\begin{figure}
\begin{center}
  \includegraphics[width=\linewidth]{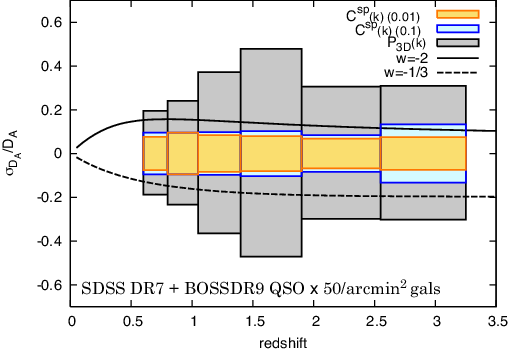} 
  \caption{
    Expected accuracies on the angular diameter distance measurements with
    the cross-correlation BAO analysis as in Table~\ref{tab:forecasts}. 
    We show the results for the photo-$z$ accuracies of both
    $\lambda_z = 0.01$ and $0.1$, denoted as $C_{\rm sp}(k) (0.01)$ or
    $C_{\rm sp}(k) (0.1)$, respectively. The outermost boxes show the
    expected accuracies when the auto-correlation power spectrum of
    the spectroscopic quasar catalogue is used. The spectroscopic
    sample is divided into six subsamples with their spectroscopic
    redshifts. The solid and dashed curves show the changes in the
    angular diameter distance when the dark energy equation of state
    ($w$) is changed to $-2$ and $-1/3$ from $w=-1$ (the cosmological
    constant). 
    \label{fig:daerror}
  }
\end{center}
\end{figure}

Table~\ref{tab:forecasts} and Fig.~\ref{fig:daerror} show an expected
accuracy of the angular diameter distance measurement in each redshift
bin via the cross-correlation method. The cross-correlation method
allows for an improvement in the geometrical test compared to the
three-dimensional auto-power spectrum analysis, by reducing the shot
noise contamination. For the SDSS/BOSS spectroscopic quasar
catalogues, the cross-correlation method improves the fractional
accuracy to better than 10\% in each redshift bin, if the photometric
galaxy survey has a full overlap with the SDSS/BOSS footprints and if we
can select adequate galaxy samples whose photo-$z$ errors are better
than $\lambda_z=0.1-0.2$ (see equation \ref{eq:photoz_accuracy}). Also, an
advantage of this method is to determine the angular diameter distance
up to a high redshift of $z\simeq 3$, where the cosmic expansion is
well in the decelerating expansion phase.
%

%
\section{Summary}
\label{sec:summary}
%
In this paper, we have studied how the cross-correlation between a
spectroscopic and a photometric sample can be used for the
two-dimensional BAO measurement. We have shown that, with the aid of
the spectroscopic sample, the cross-correlation preserves the BAO
feature in the probed transverse scales, even for the projection over
different redshifts such as $\Delta z\simeq 1$, while the angular
(cross)-correlation suffers from a smearing of the BAO feature due to
unavoidable photo-$z$ errors that cause a mixing of the different
physical scales in a particular angular scale (see Fig.~\ref{fig:cl}).
There are several notable advantages of this method. First, the
cross-correlation significantly reduces the shot noise
contamination in the measurement. Secondly, any statistical or
systematic (catastrophic) photo-$z$ errors affect only the overall
normalization of the cross-correlation function, and do not change the
shape of the power spectrum.

The cross-correlation method can be useful, if the spectroscopic
sample has a wide coverage of redshift, but does not have a
sufficiently high number density for the BAO measurement via the
autocorrelation analysis. As a specific example, we have considered
the SDSS/BOSS spectroscopic quasar sample to estimate the feasibility
of the cross-correlation method, motivated by the fact that wide-area
imaging surveys, such as the Subaru HSC Survey and {\it Euclid}, overlap 
with the SDSS/BOSS survey footprints.
Here the SDSS/BOSS quasar sample has a wide
redshift coverage of $0<z\simlt 4$ and wide area coverage of about
10000~deg$^2$, but has too small number density of $\sim
10^2~$deg$^{-2}$ per unit redshift interval to implement the BAO
measurement via the autocorrelation analysis. On the other hand, the
planned imaging surveys likely provide a much denser sampling of
galaxies such as $10^3-10^5$~deg$^{-2}$ over the redshift range. We
have shown that the cross-correlation allows a more accurate BAO
measurement over $0.7<z\simlt 3$ than in the autocorrelation of the
spectroscopic sample or the angular power spectrum of the photometric
galaxies (see Figs.~\ref{fig:delta_chisq} and \ref{fig:daerror} and
Table~\ref{tab:forecasts}), if the photometric redshift is reasonably
good, $10-20$\% level in the fractional accuracy, in order not to have
a severe dilution in the measured cross-correlation.  As shown in
Fig.~\ref{fig:daerror}, the better photo-$z$ accuracy of
$\sigma_z/(1+\bar{z}_{\rm s}) = 1\%$ does not improve constraints on $D_{\rm A}$
significantly compared to the $10\%$ photo-$z$ accuracy.  Hence the
$10-20$\% of the photo-$z$ accuracy is sufficient for the
cross-correlation BAO study, which can easily be achieved for the
current and upcoming multi-band photometric galaxy surveys.

The expected accuracy of the angular distance measurement in
Fig.~\ref{fig:daerror} is from both the BAO feature and the broad
shape of the power spectrum.  The projected cross-correlation allows
us to measure the shape of the {\em three-dimensional} power spectrum
(see Eq.~(\ref{eq:ck_approx})), although the overall normalization is
affected by photo-$z$ errors. Hence, the method can also be used to
constrain the tilt and running index of the primordial power
spectrum. Also, as an ultimate possibility, the cross-correlation
method may enable to use the observed radius of dark matter haloes in
the projected distance. If we have a good knowledge on the virial
radius of dark matter haloes as well as have a good estimator of halo
masses, to observe the virial radius can be used to infer the angular
diameter distance. This is relevant for cluster-shear weak lensing,
which probes the halo and dark matter cross-correlation
\citep[][]{OguriTakada:11}. Given that the clusters have follow-up
spectroscopic redshifts, we can expect a high-precision measurement of
the halo-matter cross-correlation at small scales down to a few Mpc,
which correspond to the virial radii of massive haloes. Thus the
virial radius may serve as another standard ruler that can lead to
even higher-precision measurements of the angular diameter
distances. This is an interesting possibility and may worth exploring
further.

We note that the cross-correlation technique developed in this paper
can also be used to constrain the primordial non-Gaussianity via
measurements of the largest-scale cross-power spectra
\citep{Dalaletal:08,McDonald:08,Slosaretal:08,Taruyaetal:08}. Again,
by measuring the cross-correlation as a function of the transverse
comoving separation, we can avoid the smearing effect due to the
projection which reduces the enhanced power at the largest scales of
$k\simlt 0.01 h/{\rm Mpc}$.

Spectroscopic observations of quasars, or more generally bright, rare
galaxies, are relatively inexpensive in terms of the observation time
needed for a given telescope. Such objects are also very interesting
subjects for astronomical studies. The method developed in this paper
can add a cosmological science case when combined with wide-area
imaging surveys that have an overlap with the spectroscopic
survey. The method is useful when designing joint spectroscopic and
photometric surveys including a science case of the two-dimensional
BAO analysis.

\bigskip
 
\section*{Acknowledgements}
We thank Chiaki Hikage for useful discussion and valuable comments.
We are grateful to the Atsushi Taruya for the use of publicly
available RegPTFast code. This work is supported in part by the FIRST
program `Subaru Measurements of Images and Redshifts (SuMIRe)',
CSTP, Japan, World Premier International Research Center Initiative
(WPI Initiative), MEXT, Japan, JSPS Core-to-Core Program
`International Research Network for Dark Energy', by Grant-in-Aid
for Scientific Research on Priority Areas No. 467 `Probing the Dark
Energy through an Extremely Wide \& Deep Survey with Subaru
Telescope' and by Grant-in-Aid for Scientific Research from the JSPS
Promotion of Science (23740161 and 23340061).


\end{document}